\documentclass[aps,twocolumn,showpacs,showkeys,amsmath,amssymb]{revtex4}
\usepackage{graphicx}
\input epsf

\newcommand{\V}[1]{\mathbf{#1}} 
\newcommand{\T}[1]{\texttt{#1}} 
\newcommand\Alfven{Alfv\'en }
\newcommand{\figref}[1]{Figure~\ref{#1}}
\newcommand{\secref}[1]{Sec~\ref{#1}}
\renewcommand{\eqref}[1]{equation~(\ref{#1})}

\begin{document}

\title{Inertial range turbulence in kinetic plasmas}

\author{Gregory G. Howes}
\affiliation{601 Campbell Hall, Department of Astronomy, University of
California, Berkeley, CA 94720, USA.}

\date{\today}

\begin{abstract}
The transfer of turbulent energy through an inertial range from the
driving scale to dissipative scales in a kinetic plasma followed by
the conversion of this energy into heat is a fundamental plasma
physics process. A theoretical foundation for the study of this
process is constructed, but the details of the kinetic cascade are not
well understood.  Several important properties are identified: (a) the
conservation of a generalized energy by the cascade; (b) the need for
collisions to increase entropy and realize irreversible plasma
heating; and (c) the key role played by the entropy cascade---a dual
cascade of energy to small scales in both physical and velocity
space---to convert ultimately the turbulent energy into heat. A
strategy for nonlinear numerical simulations of kinetic turbulence is
outlined. Initial numerical results are consistent with the operation
of the entropy cascade. Inertial range turbulence arises in a broad
range of space and astrophysical plasmas and may play an important
role in the thermalization of fusion energy in burning plasmas.
\end{abstract}

\pacs{52.30.Gz---52.35.Ra}
\keywords{Gyrokinetics---plasma turbulence}

\maketitle

\section{Introduction}
\label{}

In a warm, magnetized kinetic plasma, the dissipation of low-frequency
turbulent fluctuations by kinetic mechanisms occurs most strongly at
length scales of order the plasma particle Larmor radii. If turbulence
is driven in the plasma at a much larger scale, an inertial range
develops in which the injected energy is cascaded by nonlinear
interactions to the small scales at which the turbulence can be
effectively dissipated. This transfer of turbulent energy through an
inertial range from the driving scale to dissipative scales followed
by the conversion of this energy into heat is a fundamental kinetic
plasma physics process. In this paper, I lay out a theoretical
foundation on which to build our knowledge of \emph{inertial range
turbulence in kinetic plasmas} and outline a practical strategy for
the nonlinear numerical modeling of this process.

The study of turbulence in kinetic plasmas is not a new endeavor; in
magnetic fusion community, understanding the effect of turbulence on
the transport properties of the plasma is one of the primary goals of
the numerical modeling effort. But the turbulence most often
investigated in laboratory plasmas is driven by microinstabilities
arising from gradients in the background plasma---for example, ion
temperature gradient (ITG) and electron temperature gradient (ETG)
instabilities
\cite{Dimits:1996,Dorland:2000,Jenko:2000,Jenko:2001a,Rogers:2000,
Jenko:2001b,Jenko:2002,Candy:2004,Parker:2004}. These instabilities
inject energy into the plasma at the scale of the particle Larmor
radius; the resulting turbulence does not develop an inertial range
because the driving and kinetic dissipation scales are of roughly the
same order. Inertial range turbulence in kinetic plasmas, therefore,
represents a fundamentally different process than the kinetic
turbulence traditionally studied by the fusion community. 

The turbulence in most astrophysical contexts, on the other hand, is
typically driven at scales much larger than the Larmor radius and
gives rise to a turbulent cascade through an inertial range to the
dissipative scales at which the energy of turbulent fluctuations is
ultimately converted into particle thermal energy.  Before
constructing the detailed theoretical foundation for the study of this
process, presented in \secref{sec:theory}, I will provide here a
simplified blueprint of the physical mechanisms guiding the flow of
energy. It is important to emphasize that many of the mechanisms
described below are not well understood.  Here I aim only to paint the
overall picture in broad strokes, deferring discussion of the many
uncertainties to \secref{sec:theory}.

At a scale larger than the particle mean free path---a collisional
scale for which magnetohydrodynamics (MHD) provides an adequate
description---the plasma is stirred by some external mechanism,
driving an assortment of MHD Alfv\'en, fast, slow, and entropy mode
fluctuations in the plasma.  At this driving scale, these modes are
undamped; thus, a turbulent cascade develops nonlinearly to transfer
the fluctuation energy to smaller scales. The compressive modes become
damped as the cascade reaches scales of order or smaller than the
collisional mean free path, but the Alfv\'enic cascade continues
undamped down to the scale of the ion Larmor radius.  At this kinetic
scale, the electromagnetic fluctuations may be damped collisionlessly
by the Landau resonance with the ions. In the absence of collisions,
this process conserves a generalized energy---the free energy removed
from the electromagnetic fluctuations generates nonthermal structure
in velocity space of the ion distribution function. The remaining
electromagnetic fluctuation energy continues to cascade below the
scale of the ion Larmor radius as a kinetic
\Alfven wave cascade.  Upon reaching the scale of the electron Larmor
radius, the electromagnetic fluctuations of the kinetic \Alfven wave
cascade are completely damped via the Landau resonance with the
electrons; again, a generalized energy is conserved in this process,
leading to the creation of nonthermal structure in velocity space of
the electrons. But the damping of the electromagnetic fluctuations and
consequent generation of structure in velocity space does not
correspond to heating; irreversible heating requires an increase in
entropy that can only be achieved by collisions. The thermalization of
the turbulent energy by collisions is ultimately achieved thanks to a
cascade to small scales in velocity space of the particle distribution
functions---an \emph{entropy cascade}.  The entropy cascade drives the
distribution function structure in velocity space to scales small
enough that even weak collisions are sufficient to smooth out that
structure towards the Maxwellian, causing entropy to increase---this
is the final step in the conversion of the energy of the turbulent
fluctuations to thermal energy of the plasma particles.

The entire process described above is the kinetic generalization of
the familiar cascade of energy in a fluid turbulent system; this
fundamental kinetic plasma physics process, encompassing the dual
cascade in both physical and velocity space, is referred to as the
\emph{kinetic cascade} of the generalized energy
\cite{Schekochihin:2007}. Neither the detailed interactions of the
kinetic cascade nor its implications for the flow of energy in
turbulent systems are well understood; a theoretical foundation for
its study is layed out in \secref{sec:theory}. Nonlinear numerical
simulations of the kinetic cascade occurring in inertial range
turbulence are expected to play a leading role in shedding light on
the complex physical mechanisms involved---a strategy for the
numerical approach to this problem is presented in
\secref{sec:numerics} and initial numerical results are discussed in
\secref{sec:results}. Astrophysical and laboratory environments in
which the kinetic cascade of inertial range turbulence may play an
important role are identified in \secref{sec:app}.

\section{Theoretical Foundation}
\label{sec:theory}
Although inertial range turbulence plays an important role in the
evolution of many kinetic plasmas, it has not yet attracted great
attention within the scientific community. The physics underlying this
intersection of kinetic plasmas and turbulence is fascinating yet
remains poorly understood. This paper aims to construct a foundation
upon which a thorough knowledge of the varied mechanisms involved may
be built. I present here a view of the overall context in which the
kinetic cascade operates and aim to emphasize the uncertainty
underlying some aspects of this view.

I begin with the assumption that the turbulence is driven at a scale
$L$ much larger than the ion collisional mean free path, $L \gg
\lambda_{\mbox{mfp}i}$. The dynamics at the driving scale is therefore
collisionally dominated, so the single fluid magnetohydrodynamic (MHD)
theory provides an adequate description of the plasma dynamics at all
parallel scales $l_\parallel$ satisfying $l_\parallel \gg
\lambda_{\mbox{mfp}i}$ (where $\perp$ and $\parallel$ in this paper 
are relative to the direction of the local mean magnetic
field)\footnote{The collisional mean free path is compared to the
parallel scale of the turbulence because particle motion perpendicular
to the magnetic field is restricted by the particle Larmor motion;
here it is assumed that the ion Larmor radius $\rho_i \ll
\lambda_{\mbox{mfp}i}$.}.  The assumption $L \gg
\lambda_{\mbox{mfp}i}$ will not be true for all environments in which
inertial range, kinetic turbulence arises; the implications of
relaxing this assumption are discussed in
\secref{sec:other}.  Based on the fact that the damping of all linear
wave modes at the MHD scales $l_\parallel \gg
\lambda_{\mbox{mfp}i}$ is negligible, the damping of the turbulent
fluctuations is likewise expected to be negligible; an inertial range
will then develop to cascade the turbulent energy via nonlinear
interactions to ever smaller scales. This cascade will continue down
to the transition from fluid to kinetic behavior at a scale
$l_\parallel \sim \lambda_{\mbox{mfp}i}$. At this point, a fluid
treatment of the turbulence is no longer adequate---instead a kinetic theory
must be used to describe the dynamics. The nature of the
kinetic continuation of the inertial range to scales $l_\parallel <
\lambda_{\mbox{mfp}i}$ depends critically on the character of the MHD
turbulent fluctuations reaching this transition.

Here I review five key concepts underpinning the theory of MHD
turbulence before applying them to construct a model for the fluid
portion of the turbulent inertial range. 
\begin{enumerate}
\item The Kolmogorov Hypothesis: the nonlinear energy transfer 
is constant through all scales within the inertial range, defined as
the extent of scales influenced neither by the energy injection
mechanism nor by dissipation; at a given scale, the energy transfer is
spectrally local, with a rate determined only by the turbulent
conditions at that scale \cite{Kolmogorov:1941}.
\item The Kraichnan Hypothesis: even in a magnetized plasma with no mean
field, the magnetic field of large-scale fluctuations behaves
effectively as a mean field for fluctuations on smaller scales
\cite{Kraichnan:1965}.
\item Colliding Wavepackets: as is apparent when the equations 
of incompressible MHD are cast into Els\"asser form, only \Alfven wave
packets traveling in opposite directions along the mean magnetic
field interact nonlinearly
\cite{Iroshnikov:1963,Kraichnan:1965}.
\item Critical Balance: as strong MHD turbulence cascades to
small scale, it maintains a state of balance between the (parallel)
linear propagation and (perpendicular) nonlinear interaction
timescales \cite{Higdon:1984a,Goldreich:1995}.
\item Decoupling of Fast Wave Dynamics: theory and simulations of 
compressible MHD turbulence suggest that the isotropic cascade of fast
MHD wave turbulence decouples from the dynamics of the Alfv\'en, slow,
and entropy modes \cite{Lithwick:2001,Cho:2003}.
\end{enumerate}

\subsection{Weak MHD Turbulence}
Adopting the Kolmogorov and Kraichnan hypotheses, the interaction of
oppositely directed, small-amplitude \Alfven wave packets along the
mean (or large-scale) magnetic field in incompressible MHD is
considered. If it is assumed (a) that the nonlinear interaction
between two colliding wavepackets is weak---and therefore that a
single wave packet must undergo many collisions before it has been
distorted enough to have effectively transferred its energy to smaller
scale---and (b) that the energy is transferred to smaller scales
\emph{isotropically}, an inertial range one-dimensional energy spectrum 
that scales with wavenumber as $k^{-3/2}$ is predicted.  A spectrum
with this scaling is often referred to as the Iroshnikov-Kraichnan
spectrum \cite{Iroshnikov:1963,Kraichnan:1965}. 

However, evidence from laboratory plasmas
\cite{Robinson:1971,Zweben:1979,Montgomery:1981}, the solar wind
\cite{Belcher:1971} and numerical simulations
\cite{Shebalin:1983,Oughton:1994} suggests that the MHD turbulent cascade 
is not isotropic but preferentially transfers energy to small scales
perpendicular to the mean magnetic field.  Taking the observed
anisotropy into account, anisotropic theories of MHD turbulence have
been proposed and refined
\cite{Montgomery:1981,Montgomery:1982,Shebalin:1983,Higdon:1984a,
Sridhar:1994,Montgomery:1995,Ng:1996,Goldreich:1997}, and have led to
the emergence of a mature theory for weak incompressible MHD
turbulence \cite{Ng:1997,Galtier:2000,Lithwick:2003}. The predicted
one-dimensional energy spectrum for weak incompressible MHD turbulence
scales as $k_\perp^{-2}$; the cascade is anisotropic, with all energy
cascading to higher perpendicular wavenumbers and none to higher
parallel wavenumbers. Unlike the isotropic Iroshnikov-Kraichnan
cascade, in which the nonlinear interactions weaken as the turbulence
progresses to smaller scale, this anisotropic theory predicts that, as
the turbulence cascades to higher perpendicular wavenumber, the
nonlinear interactions strengthen, leading eventually to the
inevitable violation of the assumption of weak nonlinear interactions
\cite{Sridhar:1994,Goldreich:1997,Galtier:2000,Lithwick:2003}. 
Thus arises the most important implication of the weak turbulence theory:
given a cascade through a sufficiently broad range of scales, weak
incompressible MHD turbulence will inevitably transition to a state of
strong turbulence.

\subsection{Strong MHD Turbulence}
Below the perpendicular scale where the perturbative treatment fails,
the resulting state of strong incompressible MHD turbulence is treated
phenomenologically by assuming that the turbulence maintains a state
of critical balance, $\omega \sim \omega_{nl}$, as it cascades to
smaller scales. For a linear \Alfven wave frequency, $\omega=
k_\parallel v_A$, and a nonlinear frequency determined by the
perpendicular dynamics, $\omega_{nl} \sim k_\perp v_\perp$, theory
predicts a one-dimensional kinetic energy spectrum $E_k(k_\perp)
\propto k_\perp^{-5/3}$ and a scale-dependent anisotropy $k_\parallel
\propto k_\perp^{2/3}$. Although numerical simulations of strong MHD
turbulence with a dynamically strong mean field appear to to support
the scale-dependent anisotropy $k_\parallel
\propto k_\perp^{2/3}$ \cite{Cho:2000,Maron:2001}, these same
simulations routinely produce one-dimensional energy spectra that appear to
scale as $ k_\perp^{-3/2}$ rather than  $k_\perp^{-5/3}$
\cite{Maron:2001,Muller:2003b}. A modification of the theory to take
into account a proposed polarization alignment \cite{Boldyrev:2005}
may shed light on this discrepancy, but the matter of strong MHD
turbulence remains controversial. It is also worthwhile noting here
that the predicted transition from weak to strong MHD turbulence has
only recently been numerically reproduced \cite{Perez:2008}.

In spite of the uncertainties in the theory for the strong incompressible
MHD turbulence, the combination of key concepts (1)--(4) in
\secref{sec:theory} leads to the following general 
prediction: in any magnetized, incompressible plasma, whether or not a
strong mean field exists at the large scale, at sufficiently small
scales a state of strong MHD turbulence will arise. Therefore, given
the assumption of sufficiently large driving scale $L \gg
\lambda_{\mbox{mfp}i}$, at the transition
from fluid to kinetic behavior $l_\parallel \sim
\lambda_{\mbox{mfp}i}$, the turbulent fluctuations will be dominantly perpendicular with
$k_\parallel \ll k_\perp$ and the fluctuation amplitudes will be small
compared to the mean field, $\delta B_\perp
\ll B_0$.  The kinetic description of the
continued turbulent cascade to smaller scales is strongly influenced
by these predicted characteristics of the turbulence.

\subsection{Compressible MHD Turbulence}
The discussion thus far has been limited to incompressible MHD. In the
case of compressible MHD, an arbitrary stirring mechanism may inject
energy into the compressive modes of the plasma---the fast, slow, and
entropy modes. A weak turbulence treatment of compressible MHD for
fast and \Alfven waves suggests that only a small amount of energy is
transferred from fast waves to \Alfven waves at large $k_\parallel$
\cite{Chandran:2006}. Numerical simulations of strong turbulence in
compressible MHD demonstrate an isotropic cascade of fast waves that
scales as $k^{-7/2}$ and suggest that this cascade of fast wave
energy is decoupled from the \Alfven and slow wave cascades
\cite{Cho:2003}. In the region of wavevector space where the
the energy of the \Alfven wave cascade is concentrated, $k_\parallel
\ll k_\perp$, the fast wave frequencies $\omega \sim k_\perp \sqrt{c_s^2 + v_A^2}$,
where $c_s^2=\gamma p/(n_i m_i)$ is the sound speed, are much higher
than the \Alfven wave frequencies $\omega = k_\parallel v_A$, so the
decoupling of these cascades is not surprising. The cascade of the
slow waves and entropy modes is slaved to the \Alfven wave cascade
\cite{Maron:2001,Schekochihin:2007} and is seen to mimic
the anisotropic cascade of the \Alfven waves
\cite{Maron:2001,Cho:2003}.  Although, from these arguments, the
compressive modes of the plasma turbulence are not expected to
significantly alter the behavior of the \Alfven wave cascade as
determined by incompressible MHD turbulence theory, this topic remains
an open area of research.

\subsection{Transition to Kinetic Turbulence at  $l_\parallel \sim \lambda_{\mbox{mfp}i}$}

As the inertial range turbulent cascade enters the regime of weak
collisionality at $l_\parallel \sim \lambda_{\mbox{mfp}i}$, the nature
of the fluctuations at this transition motivates a particular choice
for the kinetic description of the dynamics of the cascade to yet
smaller scales. The theory of MHD turbulence described above predicts
the following turbulent state at the transition: isotropic fast wave
fluctuations and anisotropic Alfv\'en, slow, and entropy mode
fluctuations with $k_\parallel \ll k_\perp$. Because both the energy of the
fluctuations decreases with increasing wavenumber ($\propto k^{-7/2}$
for fast waves and $\propto k_\perp^{-5/3}$ for Alfv\'en, slow, and
entropy modes) and  the driving scale is assumed large $L \gg
\lambda_{\mbox{mfp}i}$, the amplitude of the magnetic field
fluctuations at this scale will be small compared to the local mean
field, $\delta B_\perp \ll B_0$. Although strong nonlinear
interactions dominate the turbulence---leading to the nonlinear
cascade of the energy of a wave on a timescale of order the wave
period---the small amplitude of the fluctuations suggests that it is
not unreasonable to estimate the collisionless damping of the
turbulent fluctuations using the damping rates from linear theory.
For plasmas with $\beta_i \gtrsim 1$, linear theory shows that the
kinetic wave modes corresponding to the fast and slow MHD wave modes
are significantly damped collisionlessly at scales $l_\parallel
\lesssim \lambda_{\mbox{mfp}i}$
\cite{Braginskii:1965,Barnes:1966,Foote:1979}; for lower $\beta_i$ plasmas, 
however, the damping is less vigorous.  Fast waves are also subject to
strong dissipation as they steepen into shocks.  Although the argument
that these wave modes are damped rapidly in the regime of weak
collisionality is plausible, further exploration of this matter is
required. For the remainder of this paper, I will neglect the effect
of any energy in the fast wave mode (and its kinetic continuation) on
the dynamics of the Alfv\'en, entropy, and slow modes.

In this limit $k_\parallel \ll k_\perp$ and $\delta B_\perp \ll B_0$,
the kinetic turbulent dynamics of the Alfv\'en, slow, and entropy modes
is rigorously described by a low-frequency expansion of kinetic
theory called gyrokinetics
\cite{Rutherford:1968,Taylor:1968,Catto:1978,Antonsen:1980,Catto:1981,
Frieman:1982,Dubin:1983,Hahm:1988,Brizard:1992,Howes:2006,Brizard:2007,
Schekochihin:2007}. Exploiting the timescale separation between
frequency of the turbulent fluctuations and the ion cyclotron
frequency, $\omega \ll \Omega_i$, gyrokinetics averages over the fast
cyclotron motion of charged particles in the mean magnetic field.  The
gyrokinetic approximation orders out the fast MHD wave and the
cyclotron resonance, but retains finite-Larmor-radius effects,
collisionless damping via the Landau resonance, and collisions. A
powerful result from gyrokinetic theory is the conservation of a
generalized energy $W$ according to the equation
\begin{eqnarray}
\frac{dW}{dt} & =& \frac{d}{dt} \int d^3\V{r} \left[
\sum_s \int d^3\V{v} \frac{T_{0s} \delta f_s^2}{2 F_{0s}}
+\frac{|\delta \V{B}|^2}{8 \pi} \right]  \\
&=& \int d^3\V{r} \V{J}_{a} \cdot \V{E} + \sum_s  \int d^3\V{v} 
 \int d^3\V{R}_s \frac{ T_{0s}h_s}{ F_{0s}} 
\left( \frac{ \partial h_s}{\partial t} \right)_{\mbox{c}}\nonumber
\label{eq:consw}
\end{eqnarray}
where the index $s$ denotes plasma species, $ \delta f_s$ is the
perturbation from the Maxwellian equilibrium distribution function
$F_{0s}$ with temperature $T_{0s}$, $h_s$ is the gyrocenter
distribution function defined by $\delta f_s = -q_s \phi F_{0s}/T_{0s} + h_s$
with $\phi$ as the scalar potential, $\V{J}_{a}$ is an external antenna
current driving the system, and $( \ldots )_{\mbox{c}}$ denotes the
collision operator
\cite{Hallatschek:2004,Howes:2006,Schekochihin:2007}. This relation has 
important implications for the flow of energy in kinetic turbulence,
as discussed in detail below. 

\subsection{Kinetic Turbulence at $k_\perp \rho_i \ll 1$}

The dynamics of kinetic turbulence for \emph{all} scales $l_\parallel
\lesssim \lambda_{\mbox{mfp}i}$ is  described by gyrokinetics as 
long as the frequency of fluctuations remains below the ion cyclotron
frequency \cite{Howes:2007}; the failure of the gyrokinetic theory
when $\omega \rightarrow \Omega_i$ is discussed in
\secref{sec:other}. Here I consider the properties of this turbulence
in the kinetic regime at perpendicular scales larger than the ion
Larmor radius, $l_\perp \gg \rho_i$ or $k_\perp \rho_i \ll 1$.  In this regime, Schekochihin
\emph{et al.} \cite{Schekochihin:2007} have demonstrated that the
turbulence has the following characteristics: (a) the dynamics of the
\Alfven wave cascade decouple from the compressive slow wave and
entropy mode fluctuations and are rigorously described by the
equations of reduced MHD
\cite{Strauss:1976,Montgomery:1981,Montgomery:1982}; (b) the slow and
entropy mode dynamics is governed by an ion kinetic equation and these
modes are passively mixed nonlinearly by the \Alfven waves; and (c) no
energy is transferred between the Alfv\'enic fluctuations and the
compressive fluctuations. Since the \Alfven wave dynamics are given by
reduced MHD, the \Alfven waves undergo the same anisotropic cascade
described by the theory for strong incompressible MHD turbulence and
the cascade continues undamped down to the perpendicular scale of the
ion Larmor radius $k_\perp \rho_i \sim 1$. Whether the slow and
entropy fluctuations are damped in this regime or merely passive mixed
by the \Alfven waves with little damping is uncertain at present
\cite{Schekochihin:2007}.
 
\subsection{Kinetic Turbulence at $k_\perp \rho_i \sim 1$}

At the perpendicular scale of the ion Larmor radius $k_\perp \rho_i
\sim 1$,  two effects come into play. First, the electromagnetic 
fluctuations may be damped via the Landau resonance with the ions. In
the absence of collisions or an external driving mechanism,
\eqref{eq:consw} demonstrates that the generalized energy $W$ must be
conserved during this process. Through the wave-particle interaction,
the electromagnetic fluctuations cause parallel acceleration of the
ions, transferring energy from the fields to the ions---the energy
lost by the electromagnetic fields is converted into nonthermal
structure in velocity space of the ion distribution function.

The second effect is the decoupling of the ion motion from the
electromagnetic fluctuations.  An ion samples the electromagnetic
field over the scale swept out by its fast Larmor motion. For field
fluctuations on length scales smaller than the ion Larmor radius
$k_\perp \rho_i > 1$, this leads to an averaging over the spatially
oscillating field; the net field experienced by the particle decreases
through the ring-averaging over the Larmor motion \cite{Howes:2006}.

This gradual decoupling of the ions from the fields leads to an
inherently nonlinear effect---the entropy cascade. Only recently
identified \cite{Schekochihin:2007}, the entropy cascade plays a
critical role in the inevitable thermalization of the turbulent
energy. Due to the fact that the radius of the Larmor motion for an
individual particle depends on the perpendicular kinetic velocity
$v_\perp$ of that particle, particles with high $v_\perp$ decouple
from the field more rapidly than those with low $v_\perp$. The drift
velocity of each particle perpendicular to the magnetic field is the
$\V{E} \times
\V{B}$ velocity; particles with low perpendicular kinetic velocities
$v_\perp$ feel a stronger ring-averaged electric field
$\langle\V{E}\rangle$ than particles with higher $v_\perp$, and so
drift faster. This differential $\langle\V{E}\rangle \times \V{B}$
drift leads to a coupling of structure in physical space with that in
velocity space, resulting in nonlinear phase mixing to smaller scales
in both physical and velocity space. This dual cascade is referred to
as the \emph{entropy cascade} and occurs for any species $s$ at scales
$k_\perp \rho_i \lesssim 1$. As will be explained in
\secref{sec:thermal}, this inherently nonlinear phenomenon plays a
critical role in the thermalization of turbulent energy.

\subsection{Kinetic Turbulence at $k_\perp \rho_i \gg 1$}
Any energy remaining in the electromagnetic fluctuations continues to
cascade to scales smaller than the ion Larmor radius $k_\perp \rho_i
\gg 1$ as a kinetic \Alfven wave cascade. For many parameter regimes, 
the damping of the turbulent fluctuations via the Landau resonance
with the electrons is often substantial for the entire range of scales
$k_\perp \rho_i \gtrsim 1$ \cite{Howes:2007}. Regardless of the
parameter regime, however, by the time the cascade reaches the scale
of the electron Larmor radius, $k_\perp \rho_e\sim 1$, the damping
always becomes strong enough to terminate the cascade via the Landau
resonance with the electrons. Again, in the absence of collisions, the
generalized energy $W$ is conserved and the energy lost by the fields
builds nonthermal structure in velocity space of the electron
distribution function.

\subsection{Thermalization of the Turbulent Energy}
\label{sec:thermal}
As demonstrated by Howes \emph{et al.} \cite{Howes:2006}, damping of
the electromagnetic fluctuations by the Landau resonance does not
represent heating of the plasma; collisions are required to increase
the entropy of the system and realize irreversible plasma heating. The
change in entropy $S_s$ can be expressed as
\begin{equation} 
T_{0s}\frac{d S_s}{dt} = -  \int d^3\V{v} 
 \int \frac{d^3\V{R}_s}{V} \frac{ T_{0s}h_s}{ F_{0s}} 
\left( \frac{ \partial h_s}{\partial t} \right)_{\mbox{c}}
\end{equation}
where $V$ is the plasma volume.  It can be shown that the integral is
non-positive, so the entropy always increases. Comparing to
\eqref{eq:consw}, it is clear that, in the absence of driving, the
generalized energy may only decrease due to collisions; this collision
term represents an increase in entropy and thus represents
irreversible heating of the plasma.

The action of the collision operator is to smooth out structure in
velocity space of the distribution function; the collision operator is
proportional to $\nu \partial^2 (h_s/F_{0s})/\partial \xi^2$
\cite{Schekochihin:2007} where $\xi = v_\parallel /v$ is the pitch
angle and $\nu$ is a collisional frequency. For weakly collisional
plasmas, therefore, the nonthermal structure in velocity space of
$h_s$ (created through the parallel acceleration of particles by
electromagnetic waves via the Landau resonance) must be driven to
sufficiently small scales in velocity space such that $\nu
\partial^2 (h_s/F_{0s})/\partial \xi^2 \sim \omega$; only then can 
collisions act rapidly enough to diffuse the fine scale structure in
velocity space and increase the entropy. The entropy cascade is
therefore the mechanism by which velocity-space structure is driven to
very small scales at which point the free energy contained in the
nonthermal fluctuations can be thermalized. This ultimate
thermalization of the turbulent energy marks the endpoint of the kinetic
cascade.

\subsection{Other Considerations}
\label{sec:other}
Gyrokinetic theory is only valid for dynamics with characteristic
fluctuation frequencies small compared to the ion cyclotron frequency,
$\omega \ll \Omega_i$. Since the fluctuation frequency typically
increases as the turbulence cascades to smaller scales, it is possible
to violate this condition at a scale deep into the kinetic cascade at
wavenumbers $k_\perp \rho_i \gg 1$; however, collisionless damping of
the turbulent fluctuations can slow this frequency increase along the
cascade
\cite{Howes:2007}.  Neglecting this damping provides a
conservative estimate of the perpendicular wavenumber threshold at
which the cyclotron frequency is reached; in this case, the frequency
can be estimated throughout the kinetic regime by 
\begin{equation} 
\frac{\omega}{\Omega_i} \sim \frac{( k_\perp \rho_i)^{2/3} }{\sqrt{\beta_i}} 
\left( 1 + \frac{ k_\perp \rho_i }{\sqrt{\beta_i + 2/(1+T_e/T_i)}} \right)^{2/3}
\left(\frac{\rho_i}{L}\right)^{1/3}
\end{equation}
\cite{Howes:2007}. The small scale limit of applicability of gyrokinetic theory 
must be determined on a case by case basis---see Howes \emph{et al.}
\cite{Howes:2007} for a detailed discussion of the limits of validity
of gyrokinetics in the solar wind. Astrophysical plasmas in general
support an inertial range of many orders or magnitude, so this
condition is usually satisfied to wavenumbers beyond the ultimate
termination of the kinetic cascade at $k_\perp\rho_e \sim 1$. Taking
the specific example of a plasma with $\beta_i=1$ and $T_i/T_e=1$,
this low-frequency requirement is met when $k_\perp \rho_i \ll
(L/\rho_i)^{1/4}$; for a typical astrophysical inertial range of
$L/\rho_i \sim 10^8$, this gives $k_\perp \rho_i \ll 100$, so the ion
cyclotron frequency is not reached until beyond the scale of
$k_\perp\rho_e \sim 1$ (equal to $k_\perp\rho_i \sim 40$).

Next I consider how this picture of the kinetic cascade changes if the
cascade is driven at a collisionless scale $L \lesssim
\lambda_{\mbox{mfp}i}$. Since this work is concerned with inertial range 
turbulence, it is implicitly assumed that the driving scale is much
larger than the kinetic scales, $L \gg \rho_i$. Based on the fact that
the \Alfven wave cascade, even in collisionless regimes is governed by
fluid equations for all scales $k_\perp \rho_i \ll 1$, it seems
reasonable to expect that the predictions of strong incompressible MHD
turbulence would continue to hold: an anisotropic cascade will develop
leading to an energy spectrum that scales as $k_\perp^{-5/3}$ and a
scale-dependent anisotropy $k_\parallel \propto k_\perp^{2/3}$.  If
the turbulence is driven isotropically at scale $L$, one expects that
gyrokinetic approximation will be satisfied for all scales such that
$(k_\perp L)^{1/3} \gg 1$---in other words, the specific predictions
for the kinetic cascade outlined here, based on the gyrokinetic
theory, will apply beginning at a scale several orders of magnitude
below the driving scale.

\section{Numerical Approach}
\label{sec:numerics}
The efficient numerical study of inertial range turbulence in kinetic
plasmas is best conducted using a mixture of computational
approaches. At the large, collisional scales, a fluid theory is
suitable; a hybrid scheme of fluid electrons and kinetic ions is most
efficient at intermediate scales; at the small scales where the
turbulent fluctuations are damped and their energy is converted into
heat, a kinetic scheme is required for both species. A hierarchy of
such schemes is rigorously derived from kinetic theory by Schekochihin
\emph{et al.}
\cite{Schekochihin:2007}. In this section, I describe a numerical
strategy for the study of the most fascinating aspect of
inertial range turbulence in kinetic plasmas---the transition regime
from the (reduced) MHD \Alfven wave cascade to the kinetic \Alfven
wave cascade at the scale of the ion Larmor radius.

\begin{figure}
\resizebox{3.in}{!}{\includegraphics*[angle=-90]{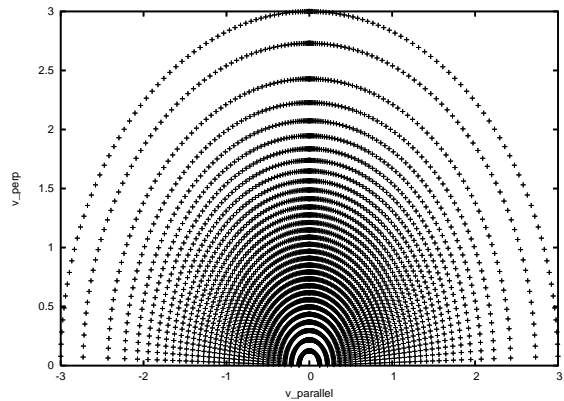}}
\caption{\label{fig:vgrid} Plot of grid used in  velocity space
with 128 pitch angles and 32 energies. Grid point locations are chosen
spectrally in a Legendre polynomial basis.}
\end{figure}

The need for a kinetic treatment of the transition regime limits the
spatial dynamic range possible in a numerical simulation.  The
strategy chosen is to model both ions and electrons gyrokinetically in
the neighborhood of $k_\perp \rho_i \sim 1$. The simulation results
reported here were obtained using \T{AstroGK}, a new gyrokinetic
simulation code designed for the numerical modeling of astrophysical
turbulence. The code is based on \T{GS2}, a mature, widely used
gyrokinetic code for the design and interpretation of laboratory
experiments in the magnetic fusion program
\cite{Kotschenreuther:1995,Dorland:2000}. \T{AstroGK} is an Eulerian 
flux tube code with periodic boundary conditions that evolves the
five-dimensional distribution function for each plasma species.  The
spatial components $x$ and $y$ are handled spectrally in Fourier
space, and the $z$-direction is handled with a compact finite
differencing scheme. Integration over the velocity-space grids is
accomplished using spectral integration by quadrature in energy and
pitch angle, while the differentiation needed by the
momentum-conserving collision operator in pitch angle is accomplished
using finite differencing.  The positions of the energy and pitch
angle grid points are determined using Legendre polynomials---an
example with 128 points in pitch angle and 32 points in energy is
shown in \figref{fig:vgrid}. The dynamics are fully electromagnetic,
evolving the scalar potential $\phi$, the parallel component of the
vector potential $A_\parallel$, and the parallel component of
perturbed magnetic field $\delta B_\parallel$ to describe the
fluctuating electromagnetic fields.  The linear terms are advanced
implicitly to avoid the need to satisfy a Courant condition for the
fast electron dynamics.  Nonlinear terms are evaluated by fast Fourier
transform in real space and the term is advanced using a 3rd-order
Adams-Bashforth scheme. A detailed description of the algorithms used
within \T{AstroGK} is given in Howes \emph{et al.} \cite{Howes:2007c}.

Because the range of scales comprising the transition regime falls
within the larger kinetic cascade, two developments are essential for
the successful modeling of this system: (a) a scheme for modeling the
inertial range energy transfer from turbulent fluctuations at scales
larger than the simulation domain; and (b) a mechanism for dissipating
the turbulence artificially to avoid the unphysical buildup of energy
at the smallest resolved scales. Both of these developments will be
described in detail and validated in a future work, but here we
provide a brief description.

Critical balance posits that the timescale for the nonlinear transfer
of energy through the inertial range is of order the linear wave
frequency at that scale in the cascade (see \secref{sec:theory}). To
simulate this process, the plasma is driven by an external antenna at
several large wavelengths comparable to the scale of simulation
domain---it is important to excite separately traveling waves moving
in both directions along the mean field in order to realize the
colliding wave packets necessary for turbulence. The amplitude and
phase of each mode determined according to a Langevin equation with a
decorrelation time of order the wave period to emulate the conditions
within the inertial range of strong MHD turbulence. The antenna drives
only a perpendicular fluctuating magnetic field to excite primarily
\Alfven waves (although at the amplitudes needed to drive strong
turbulence, fluctuations in the direction of the local magnetic field
lead to some power coupled into compressive fluctuations).

Although wave-particle interaction through the Landau resonance acts
to damp the turbulent cascade, it is often insufficient to terminate
the cascade within the resolved range of scales. To prevent a
bottleneck of energy at the smallest scales, we act on the turbulence
with a hypercollisional operator---this operator acts on the
distribution function with a non-momentum conserving collision
operator using a scale-dependent coefficient $\nu_{H
s} (k_\perp/k_{\perp \mbox{max}}) ^p_s$, where $p_s \in [4,8]$. The
heating caused by hypercollisionality is positive definite; with a
carefully chosen coefficient $\nu_{He}$, the electron
hypercollisionality is sufficient to terminate the cascade on resolved
scales.

Choosing the ideal value of $\nu_{He}$ for a given simulation is often
difficult. To avoid resorting to trail and error, an adaptive scheme
is used to choose and modify the electron hypercollisional coefficient
based on nonlinear estimations of the energy transfer frequency and of
the effective collisional (including the hypercollisional term)
damping rate at a given scale $k_{\perp H} \rho_i$. The energy
transfer frequency at $k_{\perp H} \rho_i$ is estimated by
$\omega_{nl} \simeq k_{\perp H} \sum |\delta B_\perp(k_x,k_y)|^2/8
\pi$ where the sum is over all Fourier modes $(k_x,k_y)$ within a band
of width $\Delta k_{\perp H}
\rho_i$ centered at $k_{\perp H} \rho_i$. The nonlinear damping rate
is given by the total collisional and hypercollisional heating
\cite{Howes:2006} within the same perpendicular wavenumber band
divided by the sum of the total fluctuation energy in the wave band,
$\gamma_{nl} \simeq \sum (P_{ci}+P_{ce} + P_{Hci} + P_{Hce} )/ \sum
E$. The coefficient of electron hypercollisionality $\nu_{He}$ is then
adjusted within preset bounds in order to achieve a value of
$\gamma_{nl}/\omega_{nl} \simeq 1/(2 \pi)$ within some tolerance.

Motivated by the desire to mitigate the large computational cost of
gyrokinetic simulations of kinetic turbulence, a scheme for reducing
the time to reach a statistically steady state of turbulence is
used. Initially, a small simulation is run for several eddy
turn-around times at the driving scale.  The number of cpu-hours
required to achieve a steady-state for the small problem is
moderate. The restart files from this initial run are processed to add
more spatial Fourier modes in the perpendicular direction and
finite-difference grid points in the parallel direction---this
effectively increases the spatial resolution of the simulation. The
initial values of the added Fourier modes are zero and the values at
the additional parallel grid points are determined by spline
interpolation. When the simulation is restarted, to reach a new steady
state, it only needs to run for an eddy turnaround time of the
smallest non-zero mode.  This Fourier mode expansion technique is used
recursively until the desired resolution is reached.

\section{Numerical Results}
\label{sec:results}
Here we present results from a simulation of the transition regime
from the MHD \Alfven wave cascade to the kinetic \Alfven wave cascade
for a plasma with $\beta_i=1$ and $T_i/T_e=1$.  The spatial dimensions
in the plane perpendicular to the mean field are treated
pseudospectrally on a $32\times 32$ grid; the parallel direction is
treated using a compact finite-difference scheme on 64 grid points.
Velocity space resolution uses 128 points in pitch angle and 32 points
in energy, for a total of 4096 points over the half-plane $v_\perp>0$
in velocity space; the positions of grid points are chosen
pseudospectrally for maximum accuracy when performing integrations
over velocity \cite{Howes:2007c}. The distribution of velocity space
grid points for this simulation is shown in \figref{fig:vgrid}. A
fully ionized proton and electron  plasma is specified with a realistic mass
ratio of $m_i/m_e=1836$ and both species are treated
gyrokinetically. In summary, the dimensions of the simulation are
$(n_x,n_y,n_z,n_\xi,n_E,n_s)=(32,32,64,128,32,2)$ for a total of
536,870,912 computational mesh points. The simulation used a total of
19,118 cpu-hours on Franklin, the Cray XT4 at NERSC.

A momentum conserving pitch-angle scattering collision operator is
employed for like-species collisions with $\nu_i=0.001$ and
$\nu_e=0.001$; interspecies collisions are neglected. The fully
dealiased range of perpendicular wavenumbers covers 
$k_x \rho_i \in [0.4,4.0]$ and $k_y \rho_i \in [0.4,4.0]$.

The simulation is brought to the steady-state using the Fourier mode
expansion technique described in \secref{sec:numerics}; the problem is
run using a spatial grid of $(16,16,32)$ for 1.3 outer scale periods
before expanding to $(32,32,64)$. Zero initial conditions are
specified.  The Langevin antenna, as described in
\secref{sec:numerics}, drives the six lowest wavenumber modes corresponding to 
$(k_x/k_0,k_y/k_0,k_z/k_{z0}) =
(1,0,1),(0,1,1),(-1,0,1),(1,0,-1),(0,1,-1),(-1,0,-1)$ with an
amplitude of $5.0$, a frequency of $\omega= k_\parallel v_A$ and a
decorrelation frequency of $0.8 k_\parallel v_A$.

A fixed ion hypercollisionality is used with $\nu_{Hi}=0.04$ and
$p_i=8$.  Electron hypercollisionality is chosen adaptively to
terminate the cascade as described in \secref{sec:numerics}. In the
initial (pre-expansion) run, the electron hypercollisionality is
allowed to adjust within the bounds $10 < \nu_{He} < 1200$ to meet a
dissipation criterion of $0.16<\gamma_{nl}/\omega_{nl}<0.24$ for all
modes $1.6 < k_\perp \rho_i < 2.4$; in the expanded restart, the bounds and dissipation
criterion are the same, but modes tested fall in the range $3.6 <
k_\perp \rho_i < 4.4$. The exponent for the electron
hypercollisionality is $p_e=8$.

\begin{figure}
\resizebox{3.in}{!}{\includegraphics*[angle=-90, viewport = 1.25in 0.7in 6.25in 8.1in]{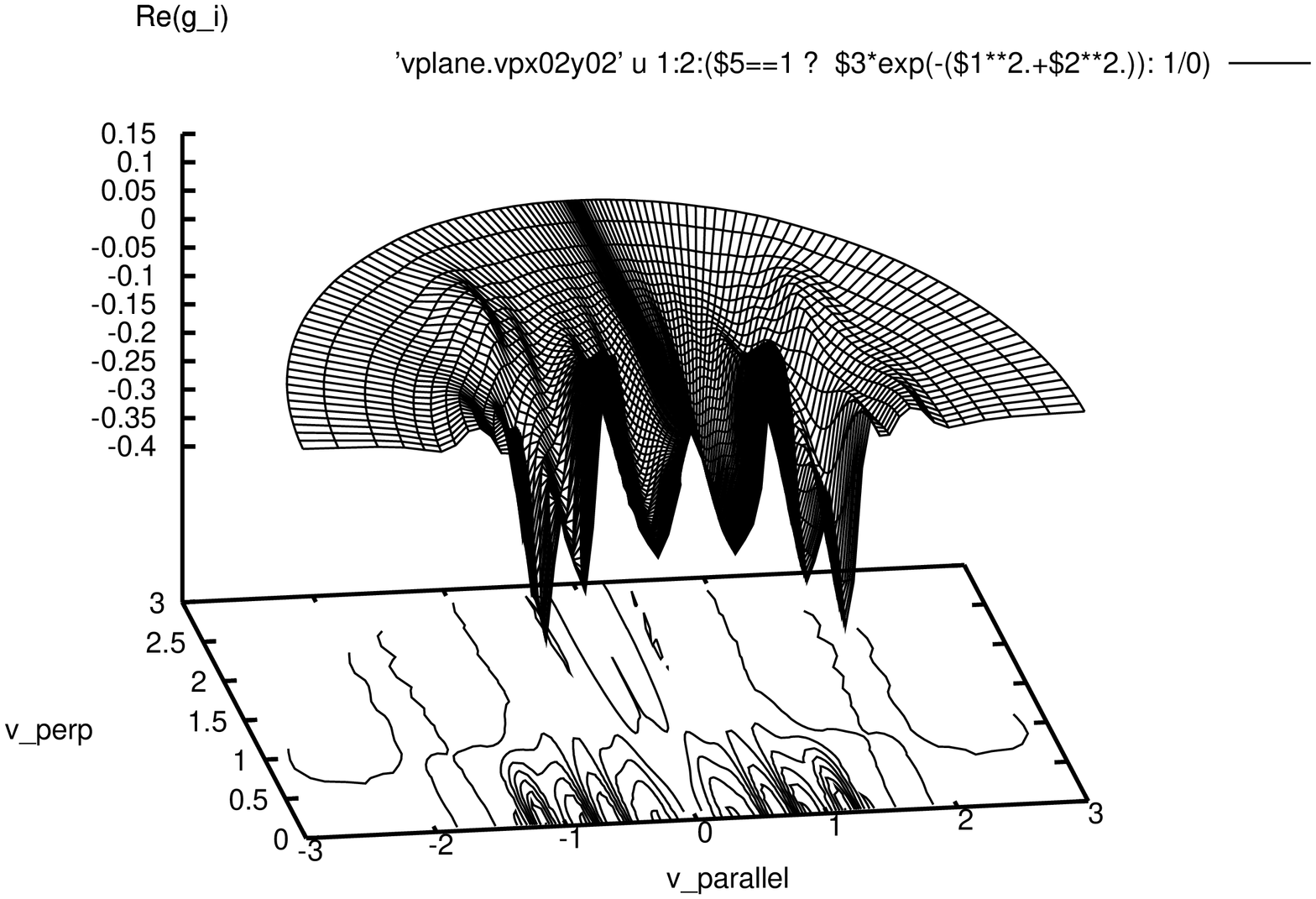}}
\resizebox{3.in}{!}{\includegraphics*[angle=-90, viewport = 1.25in 0.7in 6.25in 8.1in]{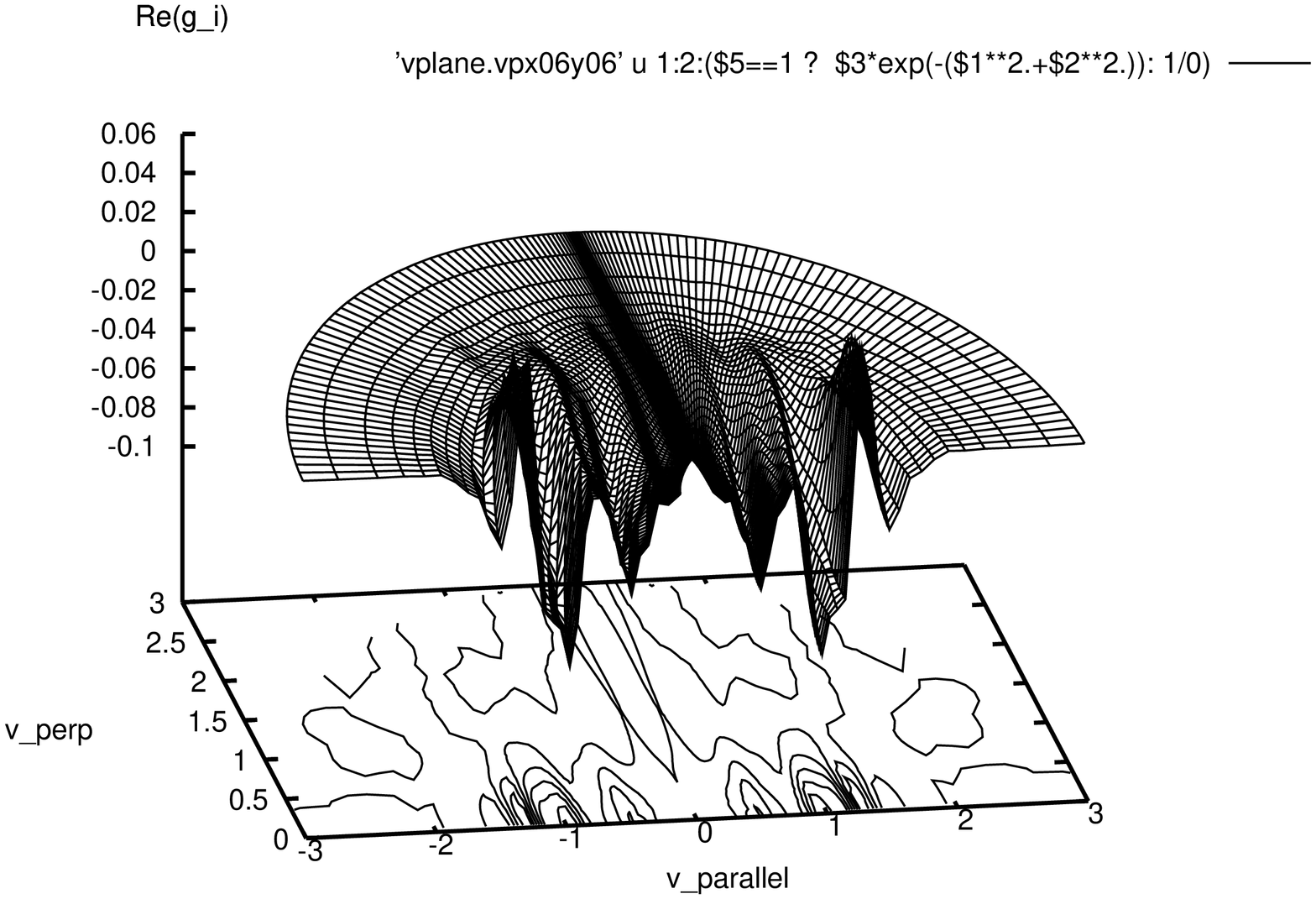}}
\caption{\label{fig:vplane} Surface plots and projected contours of the 
structure of the real part of the ion distribution function $g_s$ in
velocity space for the Fourier modes $(k_x\rho_i,
k_y\rho_i)=(0.4,0.4)$ (upper panel) and $(2.0,2.0)$ (lower panel) at
the mid-plane of the simulation domain in $z$. The lower plot, at a
wavenumber $k_\perp
\rho_i \gtrsim 1$, shows $v_\perp$ structure in velocity-space, a finding
consistent with the presence of an entropy cascade.}
\end{figure}

The high velocity-space resolution simulation presented here was set
up to investigate the entropy cascade. Presented in
\figref{fig:vplane} are surface plots and projected contours over
velocity space of the real part of the perturbed ion distribution
function evolved by
\T{AstroGK}, $g_i$, defined by
\begin{equation}
g_{i\V{k}}= h_{i\V{k}} - 
\frac{q_i \langle \phi \rangle_{\V{R}_i \V{k}} }{T_{0i}}F_{0i}
- \frac{2 v_\perp^2}{v_{ti}^2} 
\frac{\langle \delta B_{\parallel}\rangle_{\V{R}_i \V{k}}}{B_0}F_{0i}.
\end{equation}
Here the subscript $\V{k}$ denotes a Fourier coefficient and $\langle
\phi \rangle_{\V{R}_s \V{k}}$ is the gyro-average at the particle
guiding center $\V{R}_s$. The velocity space structure at the
mid-plane in the parallel direction is shown for two Fourier modes
$(k_x \rho_i, k_y \rho_i)$.  The upper plot shows the mode
$(0.4,0.4)$, corresponding to $k_\perp \rho_i \simeq 0.566$; the lower
plot $(2.0,2.0)$, $k_\perp \rho_i \simeq 2.83$. Wave-particle
interactions via the Landau resonance drive structure in $v_\parallel$
in velocity space; the nonlinear phase mixing of the entropy cascade,
on the other hand, creates structure in the $v_\perp$ direction. The
mode at $k_\perp \rho_i \simeq 0.566$ in the upper plot shows
primarily structure in $v_\parallel$ with little variation in
$v_\perp$; the lower plot, at $k_\perp \rho_i \simeq 2.83$, shows more
$v_\perp$ structure in velocity space, a finding consistent with the
action of the entropy cascade at $k_\perp \rho_i \gtrsim
1$. Velocity-space in the electron distribution function for these
same Fourier modes (not shown) is much smoother, as expected for
$k_\perp \rho_e \ll 1$.

Although the results presented here are insufficient to identify
clearly the ion entropy cascade, they represent a promising first
step. The development of more sophisticated diagnostics of velocity
space---for example, taking spectral transforms in velocity space to
yield a velocity-space power spectrum---are necessary to explore more
thoroughly the dual cascade of energy to small scales in both physical
and velocity space. As discussed in \secref{sec:thermal}, this
inherently kinetic and nonlinear mechanism plays a key role in the
ultimate conversion of turbulent fluctuation energy into heat.

\section{Applications}
\label{sec:app}
Inertial range turbulence arises across broad range of space,
astrophysical, and laboratory environments. The dissipation of the
turbulence generally occurs at scales smaller than the ion mean free
path; therefore, kinetic plasma physics determines the inevitable
plasma heating due to the dissipation of the turbulence. In this
section, I will identify a number of systems in which an understanding
of the kinetic cascade is necessary to determine the dynamics
responsible for the flow of energy and plasma heating. Identification
of the characteristic length scales for a specific system---including
the driving scale $L$, the ion mean free path $\lambda_{\mbox{mfp}i}$,
and ion Larmor radius $\rho_i$---helps to locate its place within the
theoretical framework of the kinetic cascade presented in
\secref{sec:theory}.

The most accessible turbulent astrophysical plasma is the solar wind;
the ability to make \emph{in situ} satellite measurements makes this an
invaluable environment for the study of the kinetic cascade.
Measurements of turbulence in the solar wind show a $-5/3$ magnetic
energy spectrum, interpreted as the beginning of a turbulent inertial
range, at scales $L\sim 10^{11}$ to $10^{12}$~cm
\cite{Coleman:1968,Tu:1990b,Goldstein:1995, Tu:1995,Bruno:2005}. 
The solar wind is very weakly collisional, with a mean free path of
approximately 1~AU, or $\lambda_{\mbox{mfp}i}\sim 10^{13}$~cm
\cite{Marsch:1991,Marsch:2006}. The ion Larmor radius in the solar wind 
is $\rho_i \sim 10^6$ to $10^7$~cm \cite{Leamon:1998b,Howes:2007},
yielding $\rho_i/L \sim 10^{-5}$ and giving an inertial range over
five orders of magnitude in scale. This rather small inertial range,
relative most astrophysical contexts, means the threshold at which
$\omega \rightarrow \Omega_i$ must be carefully evaluated; a more
thorough and quantitative discussion of the applicability of
gyrokinetic theory to the turbulent solar wind is contained in Howes
\emph{et al.} \cite{Howes:2007}. Although driven at a collisionless
scale $L < \lambda_{\mbox{mfp}i}$, one expects the general picture of
the kinetic cascade described here to apply, as discussed in
\secref{sec:other}. It is important to note, however, that the 
collisionality is sufficiently weak in the solar wind that the
equilibrium particle distribution functions are observed to deviate
significantly from Maxwellian\cite{Marsch:1991,Marsch:2006}.
Therefore, velocity-space anisotropy instabilities, such as the mirror
and the firehose, may play a role in the transfer of energy outside
the scope of this paper, as suggested by satellite observations
\cite{Kasper:2002,Bale:2008}.

Interstellar scintillation due to electron density fluctuations in the
interstellar medium (ISM) of our galaxy demonstrate a power law
behavior consistent with an exponent of $-5/3$ over twelve orders of
magnitude in scale, from $L \sim 10^{20}$ to $10^{8}$~cm
\cite{Armstrong:1995}. These electron density fluctuations are considered to  
be passively mixed by the \Alfven wave turbulence in the plasma
\cite{Higdon:1984a,Maron:2001,Lithwick:2001}, and thus can be used as a 
probe of the plasma turbulence. The low end of the range is consistent
with the ion Larmor radius in the ISM, $\rho_i \sim 10^8$~cm
\cite{Spangler:1990}. The mean free path in the ISM is estimated to be 
$\lambda_{\mbox{mfp}i}\sim 10^{11}$~cm. The turbulent ISM therefore
ideally fits the picture of inertial range turbulence in kinetic
plasmas described in this work.

In the accretion disk around a black hole, the magnetorotational
instability (MRI) \cite{Balbus:1991} is able to tap the energy of the
gravitational potential to drive turbulence; this turbulence leads to
enhanced angular momentum transport, allowing the plasma to accrete
onto the black hole. The MRI injects energy into the turbulent cascade
on a scale comparable to the scale height of the thick accretion disk,
$L \sim 10^{13}$~cm \cite{Narayan:2005}; this turbulence cascades down
to the scale of the ion Larmor radius $\rho_i \sim 10^4$~cm. Due to
the predicted high temperatures and low densities of the accretion
disk, the mean free path in the plasma is of order the system size,
$\lambda_{\mbox{mfp}i}\sim 10^{13}$~cm. The
radiation emitted from the hot, magnetized plasma depends on the black
hole properties and the heating of the plasma species by the kinetic
dissipation of the turbulence
\cite{Quataert:1998,Gruzinov:1998,Quataert:1999}. To understand X-ray
observational data from Chandra, the thermalization of the turbulent
energy by the kinetic cascade must be characterized.

Although the gradient-driven turbulence studied in the context of
fusion plasmas does not develop an inertial range, energetic ions in a
burning fusion plasma---arising either from Neutral Beam Injection or
as hot $\alpha$-particles generated by the fusion reaction---can drive
shear \Alfven wave instabilities at scales substantially larger than
the ion Larmor radius of the thermal plasma
\cite{Chen:1994,Campbell:2001,Gorelenkov:2003,Zonca:2006,Chen:2007}.
Experimental measurements have confirmed the generation of shear
\Alfven modes both by energetic ions \cite{Heidbrink:1993,Gorelenkov:2004} 
and by thermal ions \cite{Nazikian:2006}. The wavelength of the
\Alfven modes driven by energetic particles is comparable to the size of 
present-day experiments, so a discrete few \Alfven eigenmodes arise;
for the larger scale plasma of the International Tokamak Experimental
Reactor (ITER), a broad range of these modes may be driven, possibly
leading to the development of a turbulent cascade.  Although the
primary focus of early studies has been on the confinement of the
energetic ions in the presence of these \Alfven wave instabilities,
the kinetic cascade driven by energetic ions may prove to be an
important channel by which energy from the fusion $\alpha$-particles
is transferred to the kinetic scale of the thermal plasma and
converted into heat.

\section{Conclusions}
\label{sec:conc}

In this paper, I have presented a general picture of a fundamental
plasma physics process: the transfer of turbulent energy through an
inertial range from the driving scale to dissipative scales followed
by the conversion of this energy into heat. The details of the kinetic
cascade of the generalized energy are not yet well
understood. Numerical simulations of kinetic turbulence will play a
crucial role in illuminating the key elements of the cascade. Here I
emphasize several of these important elements:
\begin{enumerate}
\item The nonlinear cascade of energy in kinetic plasmas
preserves a generalized energy given by \eqref{eq:consw}.
\item Collisions are required to increase the entropy and thus achieve
irreversible heating of the plasma.
\item The ultimate thermalization of the turbulent energy is achieved in 
weakly collisional plasmas through an entropy cascade, a dual cascade
of energy to small scales in both physical and velocity space.
\end{enumerate}

I have outlined the numerical approach chosen to investigate
turbulence in the transition regime from the \Alfven wave to the
kinetic \Alfven wave cascade. The computationally demanding
simulations use a Langevin antenna to model driving from larger scales
in the inertial range and a hypercollisional operator to remove energy
at the smallest resolved scales. Results from a simulation with high
velocity-space resolution are consistent with the development of
$v_\perp$ structure due to the action of the entropy cascade.

Inertial range turbulence arises in kinetic plasmas across a broad
range of space and astrophysical plasmas---including the solar wind,
the interstellar medium, and accretion disks around compact
objects---and may play an important role in the thermalization of
fusion $\alpha$-particle energy in next-generation burning plasmas.

A program of nonlinear gyrokinetic simulations of kinetic, inertial
range turbulence is necessary to explore the key elements of the
kinetic cascade described here. The rich plasma physics involved in
the inherently kinetic and nonlinear nature of the predicted entropy
cascade, along with the computational intensity of the numerical work,
make this important physics problem both fun and challenging.
Additional aspects of kinetic turbulence must also be examined, such
as the effect of strong fast wave turbulence on the Alfv\'en, slow,
and entropy modes; this requires a treatment beyond the scope of
gyrokinetics.


\begin{acknowledgments}
The author thanks his close collaborators S. Cowley, W. Dorland,
G. Hammett, E. Quataert, A. Schekochihin, and T. Tatsuno for
contributing significantly to this work. This work was supported by
the Department of Energy Center for Multiscale Plasma Dynamics, Fusion
Science Center Cooperative Agreement ER54785, by the David and Lucille
Packard Foundation, the Aspen Center for Physics, and by the Franklin
Early-User Program at National Energy Research Scientific Computing Center.
\end{acknowledgments}

\appendix


\end{document}